\documentclass{PoS}

\usepackage{tikz}
\usetikzlibrary{matrix}
\usepackage{enumerate}
\usepackage{lineno}
\modulolinenumbers[5]

\def\HH{\mathsf{H}} 
\def\HHG{\mathsf{H_G}} 
\def\HHF{\mathsf{H_F}} 
\def\PPF{\mathsf{P}_{\hspace{-0.03cm}\mathrm{F}}} 
\def\PPG{\mathsf{P}_{\!\mathrm{G}}} 

\def\MM{\mathsf{M}}

\def\BH{\mathrm{BH}} 
\def\Tr{\mathrm{Tr}}

\def\pd{\partial} 
\def\eps{\varepsilon} 
\newcommand{\ket}[1]{| #1 \rangle} 

\newcommand{\avg}[1]{\langle #1 \rangle}

\title{Quasiparticle picture from the Bekenstein bound}

\ShortTitle{Quasiparticle picture from the Bekenstein bound}

\author{Giovanni Acquaviva, \speaker{Alfredo Iorio} and Martin Scholtz\\
        Faculty of Mathematics and Physics, Charles University \\
V Hole\v{s}ovi\v{c}k\'ach 2, 18000 Prague - Czech Republic\\
        E-mail: \email{gioacqua@utf.troja.mff.cuni.cz},
        \email{iorio@ipnp.troja.mff.cuni.cz},
        \email{scholtz@utf.troja.mff.cuni.cz}}

\abstract{We provide general arguments regarding the connection between low-energy theories (gravity and quantum field theory) and a hypothetical fundamental theory of quantum gravity, under the assumptions of \emph{(i)} validity of the holographic bound and \emph{(ii)} preservation of unitary evolution at the level of the fundamental theory.  In particular, the appeal to the holographic bound imposed on generic physical systems by the Bekenstein-Hawking entropy implies that both classical geometry and quantum fields propagating on it should be regarded as phenomena emergent from the dynamics of the fundamental theory.  The reshuffling of the fundamental degrees of freedom during the unitary evolution then leads to an entanglement between geometry and quantum fields.  The consequences of such scenario are considered in the context of black hole evaporation and the related information-loss issue: we provide a simplistic toy model in which an average loss of information is obtained as a consequence of the geometry-field entanglement.}

\FullConference{Corfu Summer Institute 2017 'School and Workshops on Elementary Particle Physics and Gravity'\\
		2-28 September 2017\\
		Corfu, Greece}

\begin{document}

\section{Introduction}\label{sec:intro}

The connections that have been uncovered between gravity and thermodynamics -- with the fundamental intervention of quantum effects -- have been widely regarded as clues leading to a long-sought theory of quantum gravity. In the work \cite{ACQUAVIVA2017317} we propose a new picture that might help solving some of many open issues of such a theory. Our approach starts from re-considering the bound established by Bekenstein \cite{Bekenstein1981}, who showed that the entropy $S$ of any physical system contained in a volume $V$, including the volume itself, is supposed to be bounded from above by the value of the Bekenstein-Hawking entropy $S_{\BH}$ of a black hole whose event horizon coincides with the boundary of $V$~\cite{Bousso1999}
\begin{equation}\label{eq:BekensteinBound}
	S \leq S_{\BH} = \frac{1}{4} \frac{\pd V}{\ell^2_P} \, ,
\end{equation}
where $\ell_{P}$ is the Planck length.  The generality of the original bounds for ordinary matter posited by Bekenstein is the subject of intense investigation and debates \cite{Bekenstein2014}.  Nonetheless, it is widely accepted that the bound is saturated for black holes.  Following the thermodynamic spirit, it is then natural to look for a connection between such entropy bound and the ensemble of microstates of some fundamental degrees of freedom of the system. The number of degrees of freedom $N$ of a quantum physical system is defined as the number of bits of information necessary to describe the generic state of the system.  In other words, $N$ is the logarithm of the dimension $\cal{N}$ of the Hilbert space of the quantum system. In the extreme case of a black hole ${\cal N} = e^{S_{\BH}}$.  Hence, Eq. \ref{eq:BekensteinBound}
means {\it a)} that in nature the information contained in any volume $V$ cannot exceed 1 bit every 4 Planck areas of the boundary of $V$ and {\it b)} that only in the extreme case of a black hole the hypothetical fundamental degrees of freedom are most excited (see, e.g., \cite{Bekenstein2003}). The fundamental entities characterized by such degrees of freedom cannot fully coincide with the particles customarily thought of as elementary: if it were so, the bound would be saturated with ordinary matter; moreover gravity must be included in the counting of the fundamental degrees of freedom, precisely because the saturation happens in the extreme black hole case.

Although nothing at this stage can be said about the nature of the fundamental constituents, their behaviour needs to be such that the emergent picture at our scales is that of quantum field theory (QFT) acting on a continuum classical spacetime.  However, an assumption that can be made in this context regards the character of their dynamics: does it comply with unitarity or not?  Given the central requirement of unitary evolution for the description of quantum systems at our scales, we assume such feature is preserved down to the fundamental level.  Now, the fact that both gravity and quantum fields should contribute to the counting of degrees of freedom implies the possibility of a {\it sharing} of such degrees of freedom between them.  At the same time, according to a statistical-mechanical picture, one can expect, in general, different microstates which give rise to the same classical geometry.  These configurations would yield different numbers of degrees of freedom available for the quantum fields.  Consequently, even though we assume unitary evolution on the fundamental level, the rearrangement of the fundamental degrees of freedom would lead to an entanglement between the emergent geometry and the emergent fields \cite{ACQUAVIVA2017317}.

It is worth stressing at this point that the idea that gravity is an emergent phenomenon arising from more fundamental degrees of freedom is certainly not new and goes back to Sakharov \cite{Sakharov1967,Visser2002}. Presently there exist many particular models describing how gravity could emerge and the common feature of these models is to consider some kind of underlying discrete lattice representing the mutual interactions between fundamental elements.  The fact that crystals with defects can give rise to effective non-Euclidean geometries has been employed in the cosmological ``world crystal model'' \cite{Kleinert1987}; it was proposed in \cite{VanRaamsdonk2010} how the classical properties of the space-time might emerge from the quantum entanglement between the actual fundamental degrees of freedom and a specific model along these lines has been proposed recently in \cite{Cao2016}, with the interesting possibility of recovering the ER=EPR conjecture \cite{maldacena2013,jensen2013}; finally, in quantum graphity \cite{Konopka2006,Konopka2008} the
fundamental degrees of freedom and their interactions are represented by a complete graph with dynamical structure (for more approaches see, e.g., \cite{Baez1999,Ambjorn2006,Lombard2016,Oriti2014,Rastgoo2016,Requardt2015,Trugenberger2016}).  At the same time, nonequivalent descriptions of the same underlying dynamics are a built-in characteristic of QFT \cite{Haag1992}, both in its relativistic \cite{Dirac1966} and nonrelativistic regimes \cite{Umezawa1982} (e.g., in condensed matter).  The quantum vacuum has in fact a rich structure with nonequivalent sectors or ``phases'' \cite{Milloni2013}.  Such structure is understood in QFT as due to the infinite number of degrees of freedom and/or to a nontrivial topology of the system, such as the presence of topological defects \cite{Blasone2011}.  On the mathematical level these features are the manifestation of the failure of the Stone-von Neumann theorem \cite{Neumann1931,Hall2013} that holds only for quantum mechanical systems with
finite degrees of freedom and trivial topology \cite{Bogolubov2012}. Such failure leads to the existence of different, unitarily inequivalent representations of the field algebra. That is, for a given dynamics, one should expect several different Hilbert spaces representing different phases of the system with distinct physical properties and distinct excitations deemed as elementary in the given phase \footnote{In fact, the concepts of elementary and collective excitations are interchangeable in theories where electromagnetic duality is at play \cite{Montonen1977,Castellani2016}.} \cite{Umezawa1993}, but whose general character is that of the quasiparticles of condensed matter \cite{Landau8,Landau9}.  Examples of emergent behaviours in condensed matter are the Cooper pairs of type II superconductors \cite{BSC1957,Altland2010} and the more recently discovered quasiparticles of graphene \cite{CastroNeto2009}.  In the latter case, massless Dirac quasiparticles emerge from the dynamics of electrons propagating on carbon honeycomb lattices and give rise to a continuum relativistic-like (2+1)-dimensional field theory on a pseudo-Riemannian geometry \footnote{In fact, in the case of graphene, geometries can indeed be seen as emergent \cite{Iorio2011,Iorio2015}.  Inspired by the fact that different arrangements of the carbon atoms can give rise to the same emergent spacetime geometry, in our model we take into account the possibility that the same emergent geometry can be realized through different arrangements of the fundamental degrees of freedom. These microscopic arrangements are indistinguishable at our low energies.}.  Similarly one finds examples in the context of black hole physics.  Indeed, the vacuum of a freely falling observer in Schwarzschild's spacetime can be seen, by a static observer, as a coherent state of Cooper-like pairs similar to that of a superconductor \cite{Israel1992}.  The Hawking radiation itself is related to the existence of distinct elementary excitations in the two frames (see the original derivation by Hawking \cite{Hawking1974,Hawking1976} and also \cite{Israel1976,Iorio2004}).

\section{Information loss}\label{sec:info}

It is clear from the considerations above that, even without specifying the nature (symmetries, type of interaction, etc.) of the hypothetical fundamental constituents, it is possible to obtain model-independent conclusions based on the validity of the holographic bound and the preservation of unitary evolution at the fundamental level.  At this point we can extract an important consequence for the process of black hole evaporation and the resulting information-loss issue.  In the standard scenario assuming unitary evolution, the information contained in the collapsing matter is scrambled inside the black hole, but is eventually fully released during evaporation. This paradigm of information conservation is manifested by the so-called ``Page curve'' \cite{Page1993b} (see also \cite{Harlow2016,Chakraborty2017}) which describes the complete information retrieval in the Hawking radiation at the final stage of the black hole evaporation.  On the other hand, a loss of information -- in the sense of evolution of a pure state into a mixed state -- can have two causes. The first one is that the laws of quantum theory are indeed violated in some regimes.  The second one is that only some subsystem of the universe is accessible, hence there will always be a residual entanglement of the subsystem with the inaccessible parts \cite{Wald-Unruh2017}.  We do not consider the first possibility, rather we suggest that part of the total system is always hidden in the following sense.  In the emergent picture, the probability that the fundamental degrees of freedom after the complete evaporation reorganize just like before the collapse leading to the black hole is inversely proportional to the number of their possible nonequivalent rearrangements.  Therefore, even if one demands the dynamics of the fundamental constituents to be unitary and even if the geometries before the formation of the black hole and after its evaporation are the same, the emerging quantum fields will be in general different (i.e., will live in different Hilbert spaces).  Hence, one expects that the entanglement between the geometry and the quantum fields due to the reshuffling of fundamental degrees of freedom could lead to a loss of information in the Hawking radiation \cite{ACQUAVIVA2017317}.

\section{Model of black hole evaporation}\label{sec:model}

Our goal at this point is to construct a simple kinematical model which mimics the evaporation of the black hole while taking into account the conceptual framework presented above. We consider the following idealized scenario, see \cite{ACQUAVIVA2017317}.  Initially, there is a quantum field (in an almost flat space) which collapses and eventually forms a black hole of mass $M_0$.
 The black hole starts to evaporate in a discrete way: for simplicity we assume that each emitted quantum of the field has the same energy $\eps$, so that $M_0 = N_{\max} \,\eps$ for some integer $N_{\max}$.  At the end of the evaporation, the space becomes almost flat again and the field is in the excited state with $N_{\max}$ quanta.  The formal assumptions behind such scenario are the following:
\begin{enumerate}
\item There exists a \emph{fundamental Hilbert space} $\HH$ describing the fundamental degrees of freedom of the total system (geometry and fields). Since we focus on a finite region accessible to a generic observer and big enough to contain the black hole at initial time and the emitted radiation at a later time, $\HH$ is considered finite-dimensional due to the holographic bound;

\item For a specific observer at low-energy scales, the states of $\HH$ appear as classical \emph{spatial} geometry and quantum fields propagating on it.

\item There are states in $\HH$ which represent the same classical geometry but are microscopically different.

\item During the unitary evolution there is an exchange of the number of degrees of freedom between the fields and geometry.
\end{enumerate}

In order to make connection with low-energy physics, in this model we introduce a space of classical geometries representing spatial slices of space-time containing a black hole of a given mass $M^{(a)} = a\,\eps$. That is, we introduce an orthonormal set of states
\begin{equation}\label{eq:ga state}
  \ket{ g^{(a)} }, \qquad a = 0, 1, \dots N_G - 1,
\end{equation}
where $N_G$ is therefore the number of geometries allowed in our model. For convenience, we introduce the \emph{Hilbert space of classical geometries} $\HHG$ as the linear span of the states (\ref{eq:ga state}) and
define the ``mass operator'' ${\MM}$ by
\begin{equation}
  {\MM} \ket{g^{(a)}} =  M^{(a)} \ket{g^{(a)}} \equiv \eps\,a\,\ket{g^{(a)}}.
\end{equation}
An operator of this kind should represent the possibility of measuring geometric properties of the space, such as the three-dimensional metric, as seen
by a specific observer: for simplicity, here we can restrict to cases where the geometry is determined by one macroscopic quantity (the mass, for the purpose at hand).  The assumption that the geometry of the space is a result of some coarse-graining procedure associated with a specific observer
means there is some mapping $\PPG: \HH \mapsto \HHG$ which assigns to a microscopic state in $\HH$ corresponding classical geometry
or an appropriate superposition of such geometries.  This is analogous to the {\it emergence map} recently introduced in \cite{doi:10.1142/S0218271817430131}. Similarly, we shall assume the existence of some mapping $\PPF:\HH \mapsto \HHF$ which extracts the ``field content'' of a state in $\HH$. Hence $\HHF$ can be, e.g.,\ an appropriate
Hilbert (Fock) space representing the states of the fields; more concrete definitions will depend on the
particular theory of quantum gravity. Schematically, the states of the fundamental Hilbert space $\HH$ can be interpreted as states with some classical geometry
via the mapping $\PPG$ and with some state of the quantum field via the mapping $\PPF$:
\begin{center}
\begin{tikzpicture}
  \draw (0,0) node {$\ket{\psi}\in\HH$};
  \draw[->] (-0.5,-0.3)--+(-1.5,-0.5) node[pos=0.75,above=3pt] {\scriptsize $\PPG$};
  \draw[->] (0.3,-0.3)--+(1.5,-0.5) node[pos=0.75,above=3pt] {\scriptsize $\PPF$};
  \draw (-2, -1.1) node { $\ket{g^{(a)}}\in\HHG$};
  \draw (2, -1.1) node { $\ket{\phi} \in \HHF$};
\end{tikzpicture}
\end{center}
After introducing these mappings, one can label the states in $\HH$ by the values of the coarse-grained quantities, i.e., $\ket{\psi}=\ket{g^{(a)}, \phi}$.  For simplicity we assume that any state of $\HH$ can be interpreted in such a way, although in reality
this is much more complicated: classical geometries are expected to be very special superpositions of basis
states with no classical analogues. Since we are not building
a specific model of quantum gravity, we ignore this complication. On the other hand, one can argue that among the states corresponding to
definite classical geometries one can choose a subset
of (sufficiently distinct) states which are approximately orthogonal and consider
only a subspace of $\HH$ generated by this approximately orthonormal set.

In \cite{Page1993b} Page considers a splitting of the Hilbert space representing the states of the field into ``inside'' and ``outside'' parts with respect
to the horizon of the black hole. We wish instead to implement the idea that the
geometry and its fundamental degrees of freedom must be brought into the picture, so that one should
split the fundamental space $\HH$ into a direct product of ``geometrical'' and ``field'' part. However, for our argument it is essential
to entertain the possibility that the distribution
of the microscopic degrees of freedom between the geometry and the fields is not fixed and can change during the evolution of the system.  We assume now that the fundamental Hilbert space $\HH$ can be split into a direct sum of the subspaces $T_{(i)}$,
\begin{equation}
  \HH = \bigoplus_{i=1}^{N_T} T_{(i)}, \qquad \dim \HH = N_T \, N,
\end{equation}
where each $T_{(i)}$ has a fixed dimension $N$ and consists of states with some specific distribution of the degrees of freedom between the geometry and the
fields, so that $N_T$ is the number of different available distributions. By assumption, each $T_{(i)}$ has a structure
\begin{equation}
  T_{(i)} = \HH_{\mathrm{G}}^{p_i} \otimes \HH_{\mathrm{F}}^{q_i}, \qquad p_i\,q_i = N,
\end{equation}
where $\HH_{\mathrm{G}}^{p_i}$ ($\HH_{\mathrm{F}}^{q_i}$) is a Hilbert space of dimension $p_i$ ($q_i$) representing possible microscopic states of the geometry (fields).  Considering a generic state $\ket{\psi}\in\HH$, we define its normalized projection $\ket{\psi}_{i}$ onto the subspace $T_{(i)}$.  Then the state of the field is described by the density matrix ${\rho}_{(i)}$ defined by tracing over the degrees of freedom of the geometry.  The corresponding entanglement entropy will be denoted by
\begin{equation}\label{eq:S-i}
  S_{(i)} = - \Tr_{\HH_{\mathrm{F}}^{q_i}} {\rho}_{(i)} \ln {\rho}_{(i)} .
\end{equation}
The latter represents the entanglement entropy between the geometry and the fields for a given microscopic arrangement $\ket{\psi}_{i}$ of the fundamental degrees of freedom. The expected value of the entanglement between the fields and the geometrical degrees of freedom will be
\begin{equation}\label{eq:S avg}
  \langle S \rangle = \sum_i p_{(i)} S_{(i)} ,
\end{equation}
where $p_{(i)}$ is the probability of finding the system in the state with the specific arrangement $T_{(i)}$; such arrangements are indistinguishable for the observer.

In order to explicitly compute the average entanglement entropy, we specialize to a simplified scenario which nevertheless captures the essence and consequences of the procedure: we assume that only two arrangements are possible ($N_T = 2$) and that both arrangements admit the same family of classical geometries (\ref{eq:ga state}).  Let us fix the number of degrees of freedom for each type of arrangement to $N = 1500$ and let us set
\begin{eqnarray}
  T_{(1)} & = & \HH_{\mathrm{G}}^{30} \otimes \HH_{\mathrm{F}}^{50}, \quad
                                              p_1 \times q_1 = 30 \times 50 , \nonumber \\
  T_{(2)} & = & \HH_{\mathrm{G}}^{60} \otimes \HH_{\mathrm{F}}^{25}, \quad
                                             p_2 \times q_2 = 60 \times 25 .
\end{eqnarray}
Then we have $\dim \HH = 3000$. Finally, we assume that the maximal mass $M_0$ of the black hole is split into $N_G = 30$ quanta. That is, for a black
hole of mass $M^{(a)}=a\,\eps$ there is exactly one state in $\HH_{\mathrm{G}}^{30}$ which is mapped to a state $\ket{g^{(a)}}$ by $\PPG$,
while in $\HH_{\mathrm{G}}^{60}$ there are two such states.  More details on the construction and the calculation of the entanglement entropy in this framework can be found in \cite{ACQUAVIVA2017317}.

At the beginning of the evaporation process, let the black hole have its maximal mass $M_0 = (N_G -1) \eps$ and let
there be vacuum outside the black hole.  That does not necessarily mean that the Hilbert space for the field has dimension 1, as it is in Page's case; indeed, in our model we have chosen the dimension to be either 50 or 25, depending on the microscopic arrangement.  However, since there is only one vacuum state in both arrangements, the field is disentangled from the geometry and we have $\avg{S}=0$.  Hence, our starting point coincides with the starting point of Page.  Now, as the black hole starts to evaporate,  we assume that the state in $\HH$ evolves continuously and unitarily; however we take ``snapshots'' of the system when the expected values of mass of the black hole and number of particles are respectively
\begin{equation}
  \avg{M} = (N_G-1-k) \quad {\rm and} \quad \avg{n} = k \,,
\end{equation}
where $k = 0, 1, \dots N_G-1$.
\begin{figure}
  \centering
  \includegraphics[width=1 \textwidth]{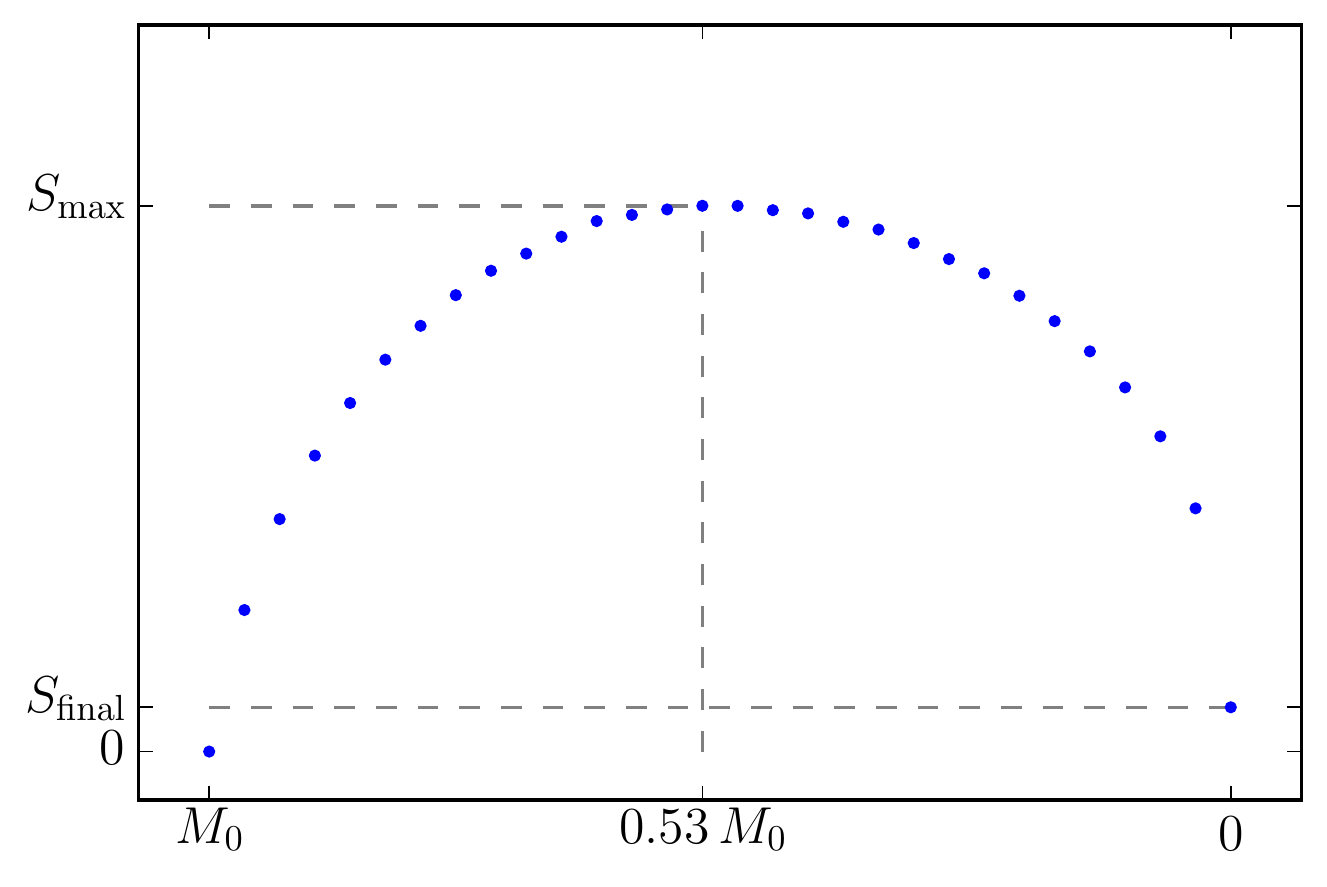}
  \caption{Entanglement entropy during evaporation in the quasi-particle picture. Entanglement entropy here is a function of the mass of the black hole, decreasing during the process of evaporation. When the evaporation starts, that is when $M=M_0$ and $\avg{S}=0$, this curve exactly corresponds to Page curve. Nonetheless, the end point corresponds to a dramatically different scenario, that is, at the end of the evaporation the entanglement entropy stays finite, due to the unavoidable entanglement between geometry and fields. This lack of pureness of the final state is due to the presence of more than one possible microscopic realization of the same emergent geometry.  One should not underestimate this effect due to the small deviation from zero we obtain in the toy model. In fact, first, even a small deviation of $\avg{S}$ from the pure state value signals a dramatic departure from the information-conserved scenario. Second, departing from the toy model, hence allowing for more microscopic realizations of the same macroscopic geometry, would in general increase the final deviation.}
  \label{fig:page-modified}
\end{figure}

An increase in $k$ corresponds to a decrease of mass of the black hole and to an increase in the number of expected particles of the field, so that, under the assumption of a dynamical evaporation process, $k$ can be taken as a discrete evolution parameter.  In Fig.\ \ref{fig:page-modified} we show the numerical result for the entanglement entropy as a function of decreasing black hole mass.  Although the curve starts at the point $(M_0,0)$, which corresponds to the same origin of the Page curve, at the final stage of the evaporation the entanglement entropy does not go back to zero.  Such deviation from a pure final state is due to the residual entanglement between geometry and field and it can be traced back to the presence of more than one possible microscopic realization of the same emergent geometry.  It is clear that allowing for more microscopic realizations of the same macroscopic geometry would in general increase the final deviation of $\avg{S}$ from the pure state value.

\section{Conclusions}\label{sec:conc}

In \cite{ACQUAVIVA2017317}, by elaborating on the fact that the number of degrees of freedom, that determine the state of a system in a compact volume $V$, is bounded from above by the Bekenstein-Hawking entropy of a black hole with horizon area $\partial V$, lead us to entertain the possibility that such fundamental degrees of freedom should describe the state of both fields {\it and} geometry contained in said volume.  The immediate consequence of such statement is that fields and geometry should be regarded as emergent phenomena at ordinary energy scales.  In this picture the particles of the Standard Model are analogous to quasiparticles, arising together with the classical background geometry from the interactions between the fundamental degrees of freedom.  We have not provided a framework for such dynamical emergence (see \cite{Cao2016,maldacena2013,jensen2013,Konopka2006,Baez1999,Ambjorn2006,Lombard2016,Oriti2014,Rastgoo2016,Requardt2015,Trugenberger2016} for some examples along these lines), and we assumed that the unitary evolution is preserved down to the fundamental level, although we are currently investigating whether such ``fundamental unitarity'' makes indeed sense \cite{vonNeumannANDBekenstein}. We then can safely conclude that the evolution inevitably leads to a {\it reshuffling} of the fundamental degrees of freedom, and this is reflected on the emergent level as an entanglement between quantum fields and geometry. In order to investigate the consequences of such scenario, we provided a kinematical framework that allows us to address the issue of information-loss in the context of black hole evaporation.  Through a simple toy model of evaporation it is shown how the entanglement between fields and geometry can lead, after the evaporation is completed, to an average loss of the initial information.  We claim that such modification of the original Page curve should be regarded as a common feature of any theory of quantum gravity in which both the spacetime geometry and the quantum fields propagating on it are emergent features of an underlying fundamental and unitary theory.

\section*{Acknowledgments}

The authors acknowledge financial support from the Czech Science Foundation (GA\v{C}R), grant no.\ 14-07983S (A.I.), and grant no.\ 17-16260Y (G.A., M.S.), and are indebted to Georgios Luke\v{s}-Gerakopoulos for many fruitful and inspirational discussions, and to Pavel Krtou\v{s} for critical remarks.

\end{document}